\documentclass[jcp,preprint,superscriptaddress,floatfix]{revtex4}
\usepackage{graphics,graphicx,dcolumn,bm,fleqn,epic,eepic,float}
\usepackage{amssymb,amsmath,multirow,rotate,color,float}
\usepackage[latin1]{inputenc}

\usepackage[dvips]{epsfig}

\begin{document}

\title{Contact Angle Hysteresis on Superhydrophobic Stripes}
\author{Alexander L. Dubov}
\affiliation{A.N.~Frumkin Institute of Physical Chemistry and Electrochemistry, Russian Academy of Sciences, 31 Leninsky
Prospect, 119071 Moscow, Russia}
\affiliation{DWI - Leibniz Institute for Interactive Materials, RWTH Aachen, Forckenbeckstr. 50, 52056 Aachen, Germany}

\author{Ahmed Mourran}
\affiliation{DWI - Leibniz Institute for Interactive Materials, RWTH Aachen, Forckenbeckstr. 50, 52056 Aachen, Germany}

\author{Martin M\"oller}
\affiliation{DWI - Leibniz Institute for Interactive Materials, RWTH Aachen, Forckenbeckstr. 50, 52056 Aachen, Germany}

\author{Olga I. Vinogradova}
\affiliation{A.N.~Frumkin Institute of Physical Chemistry and Electrochemistry, Russian Academy of Sciences, 31 Leninsky
Prospect, 119071 Moscow, Russia}
\affiliation{DWI - Leibniz Institute for Interactive Materials, RWTH Aachen, Forckenbeckstr. 50, 52056 Aachen, Germany}
\affiliation{Department of Physics, M.V.~Lomonosov Moscow State University, 119991 Moscow, Russia}


\begin{abstract}

We study experimentally and discuss quantitatively the contact angle hysteresis on striped superhydrophobic surfaces as a function of a solid fraction, $\phi_S$.   It is shown that the receding regime is determined by a longitudinal sliding motion of the deformed contact line. Despite an
anisotropy of the texture the receding contact angle remains isotropic, i.e. is practically the same in the longitudinal and transverse directions.  The cosine of the receding angle grows nonlinearly with $\phi_S$. To interpret this we develop a theoretical model, which shows that the value of the receding  angle depends both on weak defects at smooth solid areas and on the strong defects due to the elastic energy of the deformed contact line, which scales as $\phi_S^2 \ln \phi_S$. The advancing contact angle was found to be anisotropic, except in a dilute regime, and its value is shown to be determined by the rolling motion of the drop. The cosine of the longitudinal advancing angle depends linearly on $\phi_S$, but a satisfactory fit to the data can only be provided if we generalize the Cassie equation to account for weak defects. The cosine of the transverse advancing angle is much smaller and is maximized at  $\phi_S\simeq 0.5$. An explanation of its value can be obtained if we invoke an additional energy due to strong defects in this direction, which is shown to be caused by the adhesion of the drop on solid sectors and is proportional to $\phi_S^2$.  Finally, the contact angle hysteresis is found to be quite large and generally anisotropic, but it becomes isotropic when $\phi_S\leq 0.2$.

\end{abstract}
\maketitle

\section{Introduction}

Surface texture can change wetting properties in a very important way. On a surface, which is both rough and hydrophobic, the contact angle of water is often observed to be very large~\cite{quere.d:2005}. This situation is  referred to as superhydrophobic (SH). Two states are possible for a drop on such a SH solid. The first, so-called impaled state, where liquid penetrates the texture completely, is known since Wenzel~\cite{wenzel.rn:1936}.
The second idea, first expressed by Cassie and Baxter~\cite{Cassie}, assumes air is trapped by texture, so that liquid sits on the top of the asperities, being in the so-called fakir state. In such a situation the effective (`macroscopic') Cassie contact angle, which refers to a situation where the contact angle at a complex heterogeneous surface is evaluated by averaging of local angles and can be used at the scale larger than the pattern characteristic length,
is expressed as
\begin{equation}
\cos\theta^* =  -1 + \phi_{S} (1 + \cos\theta),
\label{Cassie}
\end{equation}
with $\theta$ the contact angle on the bare, smooth surface with the same chemical characteristics, and $\phi_{S}$ the solid fraction~\cite{Apparent}.

Attempts to understand SH surfaces were mostly focussed on wetting of a low density array of pillars. In this situation the Wenzel contact angle is close to Young's angle, but the hysteresis, which is usually defined as a difference between the advancing and receding contact angles, maybe very large, owing to the strong pinning of the contact line in the texture. In the Cassie state, the advancing and even receding water contact angles are outstandingly large~\cite{bico.j:1999,Oner2000}, and  hysteresis is often very low~\cite{Lafuma2003,quere.d:2005}. These remarkable (`super') properties of the SH Cassie materials have macroscopic implications in the context of self-cleaning~\cite{blossey.r:2003} and impact processes~\cite{tsai.p:2010}. However, in some recent studies of dense SH Cassie textures a contact angle hysteresis has been observed and analyzed theoretically~\cite{Whyman2008,Bormashenko2013,Extrand2002,Reyssat2009CAH,dubov.al:2012a}, similarly to reported earlier for the Wenzel state~\cite{JdG1984CAH,Semprebon2012}.  These investigations again have been mostly focussed on the isotropic arrays of pillars~\cite{Gao2006CAH,Patankar2010} and natural SH surfaces~\cite{bormashenko:2007}.

SH surfaces could also revolutionize microfluidic lab-on-a-chip systems~\cite{vinogradova2012} since the large effective slip of SH surfaces~\cite{Ybert2007,feuillebois.f:2009,vinogradova.oi:2011} can greatly lower the viscous drag~\cite{Asmolov:2011,dubov.al:2014}, and amplify electrokinetic pumping~\cite{belyaev.av:2011a,audry.mc:2010} in microfluidic devices. A physical parameter that quantifies flows is an effective slip length~\cite{vinogradova.oi:2011}, which is not directly related to a contact angle~\cite{voronov.rs:2008}. Moreover, optimization of robust transverse flows in SH devices, which is important for flow detection, droplet or particle
sorting, or passive mixing, requires highly anisotropic textures which significantly differ from those optimizing effective  slip~\cite{feuillebois.f:2010b}. In particular, it has been shown that transverse flow in thin channels is
maximized by stripes with a rather large solid fraction, where the contact angle and the effective slip are relatively small, but the Cassie state is generally stable.

These directional SH textures exhibit anisotropic wetting behavior~\cite{xia:2012,Hancock.mg:2013,Chung2007,Chen2005} and generate a contact angle hysteresis~\cite{Choi2009CAH}, which has already been employed for sorting droplets in a droplet-based microfluidic device~\cite{nilsson.ma:2011,nilsson.ma:2012}, and should be important for understanding a motion of drops on inclined anisotropic surfaces~\cite{sbragaglia.m:2014}. It is natural to suggest that on materials with a high density of anisotropic patterns, the contact angle hysteresis can be relatively large owing to an enhanced solid/liquid contact (compared to dilute micropillars). Here we discuss this quantity for striped SH Cassie surfaces.

Our paper is organized as follows. Sec.~\ref{sec_exp} contains a description of our surfaces and experimental methods. Results are discussed in Sec.~\ref{sec_results} and
we conclude in Sec.~\ref{sec_concl}.

\section{Experimental}
\label{sec_exp}
\begin{figure}[h]
  \centering
  \includegraphics[width=8cm]{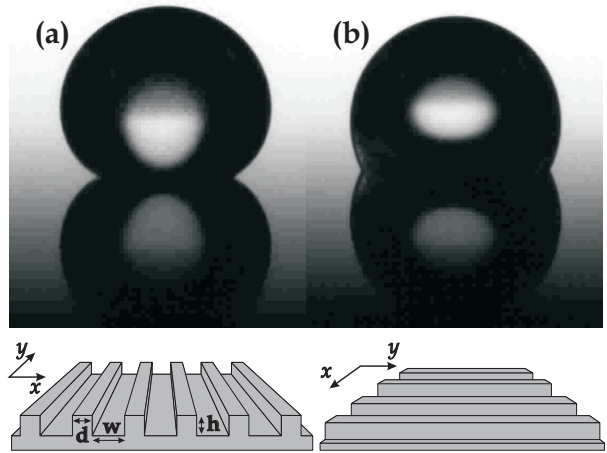}\\
  \caption{A droplet on a striped SH surface: a) transverse side view; b) longitudinal side view. }
  \label{fig_drops}
\end{figure}

SH rectangular grooves (see Figure~\ref{fig_drops}) of width $w$ separated by distance $d$ (both varied from 3 to 100~$\mu$m) and depth $h$ (5~$\mu$m) have been prepared to provide $\phi_S = d/(w+d)$ from 0.12 to 0.88 in the Cassie state. To manufacture the SH surfaces we used two methods. The silicon masters fabricated by projection photolithography have been ordered from AMO Gmbh, Germany. They have been oxidized by plasma treatment in evacuated chamber with oxygen flow 30~mL/min and plasma power 200~W (PVA TePla 100 Plasma System), and coated with a trichloro(1H,1H,2H,2H-perfluoroocty)silane (PFOTCS, 97\%, Sigma Aldrich) through the gas phase in evacuated dessicator at pressure 80~mbar for 3~hours. After removal of a bowl with PFOTCS from dessicator it was evacuated to full vacuum for 15~minutes in order to remove formed  clusters of PFOTCS from the silicon surface. Hydrophobized surfaces have been rinsed by isopropanol. These surfaces were used for all contact angle measurements. Flat silicon surfaces have been also hydrophobized similarly. Second, we fabricated intrinsically hydrophobic polydimethylsiloxane (PDMS)  replicas from silicon patterns by a common soft lithography~\cite{Xia1998}. PDMS was prepared by mixing the rubber component A with the curing agent B (both Sylgard\circledR 184 Silicone elastomer KIT, Dow Corning) 10:1 by weight, then the mixture was degassed, poured onto silicon masters and cured for 3~hours at 60$^\circ$. Afterwards cured PDMS replicas have been carefully pilled from the silicon masters and used without any further treatment. These (transparent) PDMS surfaces were used for some contact angle measurements and also for monitoring of the contact line motion.

The geometry of the patterns after hydrophobization has been validated by using the interference profilometry (WYKO NT2000, Veeco, USA) and optical microscopy (Axioplan 2 Imaging System, Zeiss, Germany), which also allowed obtaining exact values of $w$, $d$, $h$ and $\phi_S $ for each sample.

To measure the  \emph{receding} and \emph{advancing} contact angles we use the Drop Shape Analysis System (DSA100, Kr\"uss, Germany) by employing two independent methods. In the first method~\cite{drelich1996} the volume of the sessile drop oscillated in the range 7$\pm$5~$\mu$L being fed with a syringe. The advancing angle is then measured by introducing purified milli-Q water and is the largest before the drop advances. Similarly, as water is drawn out of the drop, the smallest angle possible before the drop retracts is the receding contact angle. In the second method we investigated the directed motion of the droplet of a volume ca. 4~$\mu$L driven by the syringe needle placed inside the droplet with constant (low) velocity~\cite{reyssat2010}. The angle at the front edge is the advancing, and that at the rear, the receding contact angle. Note that first method gives the maximal values of advancing contact angles, while the second method gives lowest values of receding angles. Since the measured receding contact angle is known to depend on the method used, we have verified our values with the receding angle obtained from the evaporation of water droplet (see~\cite{Bormashenko2011} for a description of this method) and found an excellent agreement with the data we present here.
Values of contact angles were obtained by averaging over $5-10$ measurements. In case of striped surfaces all contact angles were measured in two, longitudinal and transverse, eigendirections of the surface pattern. The notations we use below are summarized  in Table~\ref{table_design}. We monitor the local shape of triple contact line by using optical microscopy (Axioplan 2 Imaging System, Zeiss, Germany) and confocal microscopy (Leica TCS SP8, Germany).

\begin{table}[h]
\small
  \caption{\ Notations of different contact angles on flat and patterned surfaces}
  \label{table_design}
  \begin{tabular}{lll}
    \hline
    Type of surfaces & receding &  advancing \\
    \hline
    Flat & $\theta_{r}$ &  $\theta_{a}$ \\
    Patterned & $\theta_{r}^{\ast}$ &  $\theta_{a}^{\ast}$ \\
    \hline
  \end{tabular}
\end{table}

\section{Results and discussion}
\label{sec_results}

\subsection{Contact angles on the flat hydrophobic surface}
\label{Sec_Flat}

We begin by studying contact angles on the flat hydrophobized silicon surface. The cosines of measured contact angles are presented in Table~\ref{table_flatCAH}  and clearly show that there is some contact angle hysteresis. Note that for all measurements errors did not exceed instrumental error of 1$^\circ$. Our data are in good agreement with prior work~\cite{extrand1997}, which has been interpreted by the static friction for the contact line displacement~\cite{yaminsky2000}.


\begin{table}[h]
\small
  \caption{Contact angles of water on the flat hydrophobized surface used in the study. }
  \label{table_flatCAH}
  \begin{tabular}{lll}
    \hline
     & receding & advancing \\
    \hline
    $\theta$,$^\circ$ & 84 & 115 \\
    $\cos\theta$ & 0.1 & -0.42 \\
    \hline
  \end{tabular}
\end{table}

Measured hysteresis indicates that flat hydrophobic surface contains weak defects, likely small scale chemical heterogeneities or some roughness elements~\cite{JdG1984CAH}.

\subsection{Contact angle hysteresis on the patterned surface}
\label{Sec_Patterned}

\begin{figure}
  \centering
  \includegraphics[width=4cm]{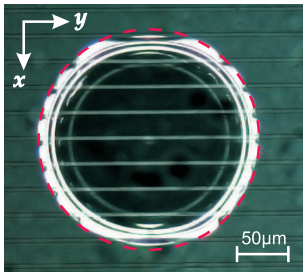}\\
 \caption{A bottom view of droplet baseline on a striped Cassie surface with $\phi_S=0.87$. }
  \label{fig_contact_base}
\end{figure}

Now, we measure contact angles on the striped superhydrophobic surface with different $\phi_S$. Figure~\ref{fig_drops} shows red a typical sessile drop on a grooved Cassie surface. We see that the effective contact angle is different in transverse and longitudinal directions. A typical (bottom) view of the droplet baseline is shown in Figure~\ref{fig_contact_base}. It can be seen that the shape of the contact line base slightly deviates from circular. Thus, unlike the homogeneous flat surface, the hydrophobic striped surface demonstrates anisotropic wetting
properties, which depend on $\phi_S$.

Below we summarize our data for receding and advancing contact angles. Taking into account the anisotropy of wetting we measure contact angles in two eigendirections.

\subsubsection{Receding Contact Angles. }

\begin{figure}
  \centering
  \includegraphics[width=8cm]{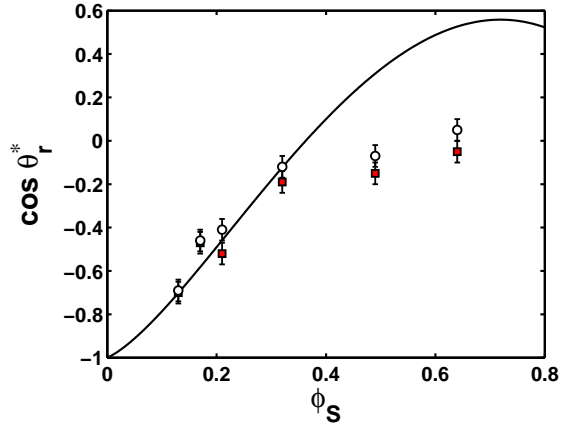}\\
  \caption{Receding contact angles for different solid-liquid fractions. Filled squares are experimental data for transverse direction, open circles - for longitudinal direction. Solid curve shows predictions of Eq.(\ref{eq_rec}).}
  \label{fig_RCA}
\end{figure}

Let us first consider the receding contact angle.  Note that it was measured only for patterns with $\phi_S<0.7$ since experiment in a high $\phi_S$ regime was affected by the Cassie-to-Wenzel transition. We leave the study of this transition for a future work. In Figure~\ref{fig_RCA} we have plotted  the experimental results for the longitudinal and transverse receding contact angles as a function of $\phi_S$.
A first striking and a counterintuitive
result is that receding angle is practically isotropic. Note, however, that observations of an isotropic receding angle for a drop on striped surfaces have been reported before~\cite{Choi2009CAH}. There have been also similar observations made for anisotropic arrays of pillars~\cite{Gauthier2013,dorrer.c:2006}. As expected, the cosine of the receding contact angle is small in the dilute regime and increases for denser patterns. However, this is the most important result emerging from this plot, the increase is nonlinear in contrast to predictions of classical Eq.(\ref{Cassie}).

\begin{figure}[h]
  \centering
  \includegraphics[width=8cm]{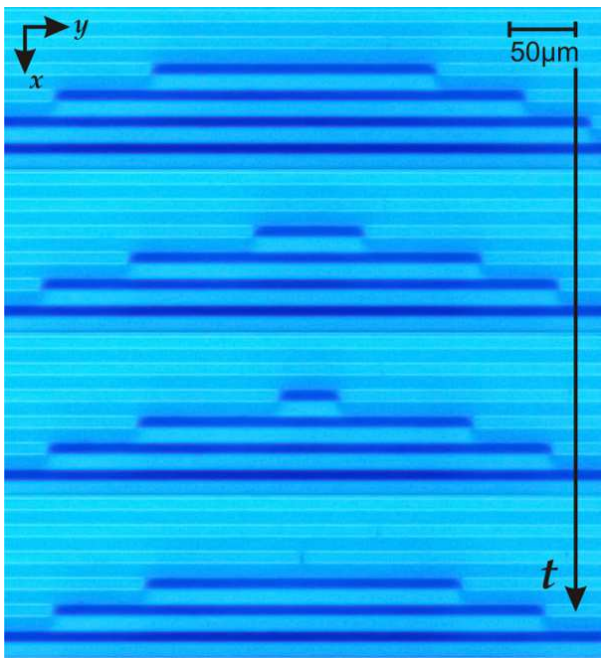}\\
  \caption{Experimental recording of the  contact line receding reflecting
 its shape and evolution in time. }
  \label{fig_receding}
\end{figure}

 To understand this isotropic behavior we explored the local motion of the contact line, and found it to be irregular. In other words, the contact line does not move smoothly, it sticks and slips.
 This is illustrated in Figure~\ref{fig_receding}, which shows a sequence of snapshots of the contact line near the edge of the drop. In the transverse direction, $x$, the line recedes in abrupt steps of the size $w+d$ and remains pinned for a period before next step. The pinned state of the contact line in transverse direction indicates, that in this situation the contact line slides
in the longitudinal direction $y$. These features are qualitatively similar to described before for stripes in the Wenzel state~\cite{degennes.pg:1985}, and is likely responsible for an observed isotropy of the receding angle.  Our results thus show that in contrast to two-dimensional surfaces, where receding is discontinuous and its isotropy just reflects the isotropy of the surface texture, anisotropic one-dimensional grooves generate an isotropic receding due to a continuous longitudinal sliding of the contact line.

\begin{figure}
  \centering
 \includegraphics[width=8cm]{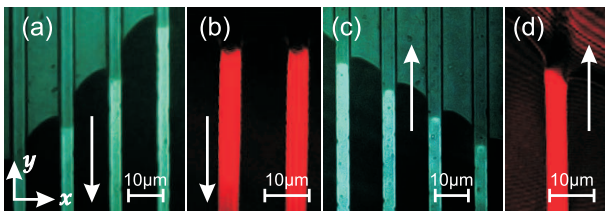}\\
  \caption{The images of the shape of the contact line  obtained by optical (a, c) and confocal (b, d) microscopy for receding (a, b) and advancing (c, d) motion. Arrows indicate the direction of motion. Dark stripes are the solid and bright stripes (cyan and red) are the gas areas of a superhydrophobic surface. }
  \label{fig_contact_line}
\end{figure}

Figure~\ref{fig_contact_line} (a) and (b) show the shape of the contact line during receding in smaller scale. It is well seen that at gas areas the contact line is bent in the direction of motion (or of a force), and is always ahead the contact line at solid areas, being pinned by edges.
We remark that this scenario is essentially different from reported for two-dimensional patterns (e.g. pillars)~\cite{Krumpfer2011}, since our continuous receding (along one-dimensional grooves) obviously cannot involve the rupture of capillary bridge.

Now we propose a simple description to explain the nonlinearity of the receding angle and its deviations from predictions of Eq.(\ref{Cassie}). The observations of the motion and shape of the receding contact line suggest arguments, based on the earlier model of hysteresis on a flat surface~\cite{JdG1984CAH}, which has been  recently adapted for isotropic SH textures~\cite{Reyssat2009CAH,dubov.al:2012a}. We first modify the Cassie equation, by substituting $\theta$ by $\theta_{r}$

\begin{equation}
\cos\theta^{\ast}_{r}= - 1 + \phi_{S} (1 + \cos\theta_{r}),
\label{Cassie_rec}
\end{equation}
The physical idea underlying this relationship is to account for weak defects that lead to a hysteresis on a flat solid areas. Eq.(\ref{Cassie_rec}) however cannot fit the data presented in Figure~\ref{fig_RCA} and contains only linear terms in $\phi_{S}$. The same remark concerns an earlier model for a receding angle on SH stripes~\cite{Choi2009CAH}.

\begin{figure}[h]
  \centering
  \includegraphics[width=4.5cm]{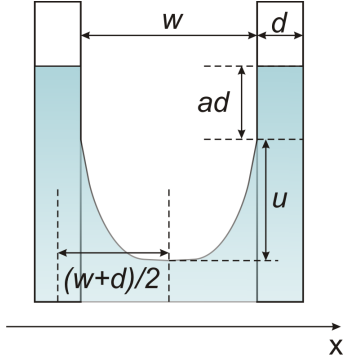}\\
  \caption{Sketch of deformed contact line}
  \label{fig_deformation}
\end{figure}

 To take into account the strong defects at the borders of stripes we evaluate the
elastic energy per unit area, $E$, similarly to suggested in prior work for an isotropic array of pillars~\cite{Reyssat2009CAH}. This model assumes that when trying to displace a liquid on such solids, the contact line largely deforms, as sketched in Figure~\ref{fig_deformation}. Following earlier ideas~\cite{JdG1984CAH,quere.d:2008} (formulated for the limit of low $\phi_S$, i.e. by assuming that $w\ll d$) we express the deformation, $u$, of the local contact line at the gas area as
$x=d\cosh(u/d) \simeq d\exp(u/d)/2.$ Maximal deformation is reached in the middle of the gas gap between two solid stripes, $x=(w+d)/2$, and can be evaluated as
\begin{equation}\label{deformation}
 u\simeq d\ln\left(\frac{w+d}{d}\right).
\end{equation}
The force on each strong defect can be written as
\begin{equation}\label{force}
    F \simeq \gamma d (2a+1)   \simeq  \frac{\gamma (2a+1) u}{ \ln(\frac{w+d}{d})},
\end{equation}
where we introduced a length of a side part of the contact line contour on the solid stripe, $a d$, (see Figure~\ref{fig_deformation}) which has been observed experimentally (see Figure~\ref{fig_contact_line}(a)), and therefore $ d (2a+1)$ is a perimeter of the pinned contact line. Force $F$ and deformation $u$ are proportional to each other, so that we can define a spring stiffness of the contact line $K=\gamma (2a+1)/\ln\left(\frac{w+d}{d}\right)$, and the energy stored in the deformation of the contact line at the gas areas is then $Ku^2/2$. By using a definition of $\phi_S$, we then derive for this energy per unit area

\begin{equation}
\frac{E}{\gamma} \simeq - \frac{2a+1}{2} \phi_S^2\ln\phi_S.
\label{eq_rec1}
\end{equation}


Now it is naturally to assume the additivity of weak and strong defects. By adding then  Eq.(\ref{eq_rec1}) to Eq.(\ref{Cassie_rec}) we finally get

\begin{equation}
\cos\theta_r^{\ast}= - 1 + \phi_{S} (1 + \cos\theta_{r})-\frac{2a+1}{2}\phi_S^2\ln\phi_S.
\label{eq_rec}
\end{equation}
This expression consists of two terms in $\phi_{S}$. The linear  term
just reflects the additivity of week defects. The non-linear (logarithmic) term originates from the shape of
the distorted line on the strong defects, and is sensitive both to its length on the solid area and the defect density.

We fitted the experimental data to Eq.(\ref{eq_rec}) taking $a$ as a fitting parameter. The theoretical curves are included in Figure~\ref{fig_RCA}, and the value $a=4.1 \pm 0.5$ was obtained from fitting. Experimental observations (Figure~\ref{fig_contact_line} (a)) suggest that $a$ should be rather of the order of 2 or 3. However, any deviations of the contact line shape on the solid stripes from the ideal rectangle used in our estimates  would increase this factor. These deviations are due to irregularities at the edge of pinning and some curvature of the contact line on solid sectors, where defects are weak. Therefore, the value of $a$ deduced from the fit seems to be physically reasonable. It is natural to suggest that the above mentioned sensitivity of the receding contact angle to the experimental procedure used is due to different value of $a$, which can be influenced by the measuring procedure.
 A general conclusion from Figure~\ref{fig_RCA} is that the theoretical  predictions  are in good agreement with experimental results for $\phi_S\leq0.35$, confirming the validity of the approach, based on elastic distortion of the contact line, for a relatively dilute anisotropic striped pattern. For larger $\phi_S$ the experimental data significantly deviate from the fit. The condition of dilute stripes is not satisfied, and the model underestimates the value of the effective receding angle. Note however that the range of applicability of Eq.(\ref{eq_rec}) is quite large and the same as reported in previous studies for a texture decorated by circular pillars~\cite{Reyssat2009CAH}. We finally remark that previous work~\cite{Reyssat2009CAH} ignored the presence of weak defects at the solid area, i.e. used $\theta$ instead of $\theta_r$ in their equation for $\cos\theta_r^{\ast}$. We have therefore tried to make a similar analysis, but were unable to provide a reasonable fit to the data for the whole range of $\phi_S$. This (together with a contribution to $a$) clearly shows the importance of week defects on smooth solid sectors in determining the behavior of receding contact angles on patterned surfaces.

Let us now turn to the advancing contact angle.

\subsubsection{Advancing contact angles. }

\begin{figure}
  \centering
  \includegraphics[width=8cm]{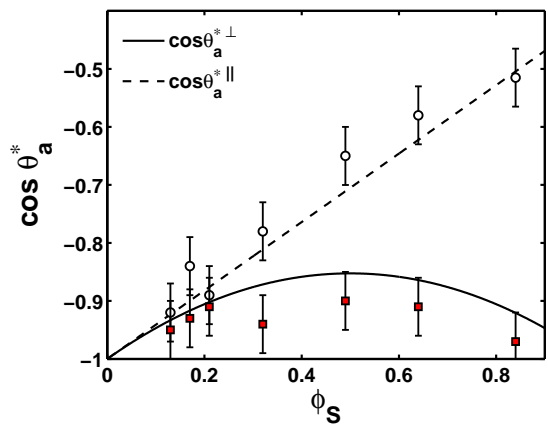}\\
  \caption{Advancing contact angles for different solid-liquid fractions. Filled squares are experimental data for transverse direction, open circles - for longitudinal direction. Dashed line shows predictions of Eq.(\ref{Cassie_adv}), solid curve are calculations made with Eq.(\ref{eq_advancing}).}
  \label{fig_ACA}
\end{figure}

The experimental results for two eigendirections are shown in Figure~\ref{fig_ACA}. The main conclusion from this plot is that the advancing angle is strongly anisotropic, except in a very dilute regime.

For the longitudinal direction the cosine of the advancing angle depends linearly on $\phi_S$. The data are well fitted by straight line, but deviate from the Cassie predictions, Eq.(\ref{Cassie}). Therefore, we try to generalize the above ideas that led to Eq.(\ref{eq_rec}) to predict

\begin{equation}
\cos\theta^{\ast}_{a}= -1 + \phi_{S} (1 + \cos\theta_{a})
\label{Cassie_adv}
\end{equation}
Theoretical curves are included in Figure~\ref{fig_ACA} and we conclude that they are in good agreement with experimental results. This suggests that the advancing longitudinal angle is hindered or restricted by weak defects, and that the contact line is not pinned at the edges of stripes. We note that an earlier model~\cite{Choi2009CAH}  predicts in this situation the validity of Eq.(\ref{Cassie}), so it does not take into account possible week defects, which are important as we have demonstrated here.

Monitoring of the shape of the transient contact line during the longitudinal advancing of the drop shows (see Figure~\ref{fig_contact_line}(c) and (d)) that the solid part of the contact line is ahead its gas sector, which is indeed not pinned. A conclusion emerging from this observation is that in contrast to receding (sliding) regime, the advancing of the drop represents rather its smooth rolling motion, which immediately explains Eq.(\ref{Cassie_adv}). Indeed, by assuming the rolling of the drop, one can speculate that a precursor contact line on solid areas, which is likely formed since their wetting by a liquid is more preferable than that of the gas parts, determines the longitudinal advancing angle.

The data obtained for a transverse direction generally show smaller $\cos \theta_a^*$, which is maximized by stripes with $\phi_S\simeq0.5$. This result immediately rules out the earlier model~\cite{Choi2009CAH}, which  considers  only a contact line deformation and predicts for this case $\cos \theta_a^* = 180°$, independently on $\phi_S$. The transverse advancing of the transient contact line represents an irregular motion due to the presence of strong defects (borders of gas/solid areas in this direction), which is however physically different from the transverse receding (or a similar scenario expected for two-dimensional SH surfaces~\cite{dubov.al:2012a}). This is because (and in contrast to the assumption of the key role of the contact line~\cite{Choi2009CAH}) for a transverse advancing on SH stripes the role of a distortion of the contact line at the gas sectors is negligibly small compared to a solid area contribution.

An explanation for a value of a transverse angle can be then obtained if we invoke additional energy due to strong defects.
The contact line of length $dy$ pins as we move the liquid, by applying a force. The line meets $\phi_{S}$ defects for a displacement $dx$, or $\phi_{S} dx$ defects  per
unit length.
 Passing each of them, an energy $dE$ is stored and then released in the liquid as the line depins~\cite{quere.d:2008}. We now make a simple suggestion that depinning happens when adhesion on the solid area $\phi_S\gamma(1+\cos\theta_a)dy$ is equal to applied force, which immediately leads to

\begin{equation}
\frac{E}{\gamma}= \phi_S^2 (1+\cos\theta_a)
\end{equation}
Inclusion of this term into Eq.(\ref{Cassie_adv}) gives

\begin{equation}
 \cos\theta_a^{\ast}=-1 + \phi_{S} (1 + \cos\theta_{a}) + \phi_S^2(1+\cos\theta_a)
 \label{eq_advancing}
\end{equation}
Thus similarly to the case of the receding angle we derived a non-linear expression, which contains two terms in $\phi_{S}$. The linear term again reflects the additivity of weak defects. However it is now corrected by the quadratic term, which reflects an additional energy of adhesion on the solid areas. Note that our result confirms the hypothesis~\cite{Bormashenko2013}, that when the advancing contact angle is governed by the solid area wetted by a droplet, it should be somehow related to $1+\cos\theta_a$. This is exactly what Eq.(\ref{eq_advancing}) predicted.

We emphasize that this is merely a simple-minded approach which we will use in an attempt to fit experimental data. In particular, we should like to note that it does not provide a correct asymptotics at $\phi_S\simeq 1$. However, the predictions of Eq.(\ref{eq_advancing}) included in Figure~\ref{fig_ACA} are in quantitative agreement with experiment, proving that our tentative model is physically correct in an extremely large range of $\phi_S$.

\subsubsection{Contact angle hysteresis.}

Finally, we consider in this paragraph the contact angle hysteresis, $\Delta \cos \theta^\ast = \cos\theta_r^\ast - \cos\theta_a^\ast$, which can be easily evaluated from the above results. By subtracting Eq.(\ref{Cassie_adv}) from Eq.(\ref{eq_rec}) we derive for a longitudinal direction

\begin{equation}\label{hysteresis_lon}
  \Delta \cos \theta^\ast = \phi_S (\cos\theta_r - \cos\theta_a)-\phi_S^2 \frac{2a+1}{2}\ln\phi_S
\end{equation}
By using Eqs.(\ref{eq_rec}) and (\ref{eq_advancing}) we get for a transverse direction:
\begin{eqnarray}\label{hysteresis_tr}
  \Delta \cos \theta^\ast &=& \phi_S (\cos\theta_r - \cos\theta_a)\nonumber\\
  &-&\phi_S^2 \left(1 + \cos\theta_{a} + \frac{2a+1}{2}\ln\phi_S \right)
\end{eqnarray}

Here the linear in $\phi_S$ terms reflect a contribution of the material itself, i.e. weak defects on the solid area, and second terms gathers all the effects due to the existence of a texture, which generates strong defects at the borders of stripes. We stress that the latter are quite large and essentially non-linear in $\phi_S$ even for dilute stripes (due to a logarithmic term), and recall that the physical origin of strong defects is essentially different for receding and advancing of the drop.

\begin{figure}
  \centering
  \includegraphics[width=8cm]{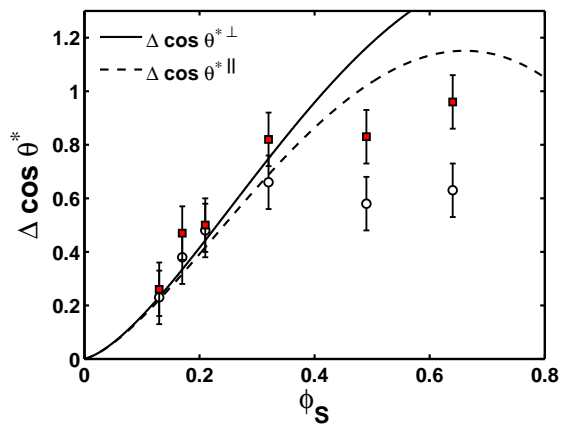}\\
  \caption{Contact angle hysteresis $\Delta \cos \theta^\ast$ for different solid liquid fractions. Filled squares are experimental data for a transverse, and open circles for a longitudinal direction. Dashed and solid curves show theoretical predictions, Eqs.(\ref{hysteresis_lon}) and (\ref{hysteresis_tr}).}
  \label{fig_CAH}
\end{figure}

We have plotted in Figure~\ref{fig_CAH} experimental and theoretical $\Delta \cos \theta^\ast$ as a function of $\phi_S$. This plot illustrates immediately that hysteresis is quite large. Both absolute value of hysteresis and its anisotropy increase with $\phi_S$. An important conclusion from our analysis is that the hysteresis becomes isotropic when $\phi_S\leq 0.2$. We emphasize that the experimental results are in a very good quantitative agreement the predictions of our simple models for $\phi_S\leq0.35$ and start to deviate from the predictions when the condition of dilute stripes is not satisfied anymore. Finally we note that in contrast to common beliefs  and our initial expectations the values of $\Delta \cos \theta^\ast$ for striped SH Cassie surfaces were found to be of the order of observed for arrays of pillars with the same area fractions~\cite{Reyssat2009CAH}.

\section{Conclusions}
\label{sec_concl}

We have measured receding and advancing contact angles of a droplet on striped Cassie surfaces with a different fraction of solid areas. Our experiment demonstrated that the receding angle is isotropic, and increases nonlinearly with $\phi_S$. The receding contact line shows irregular sliding motion. Results were found to be in a quantitative agrement with predictions of a simple theoretical model, which takes into account the role of both weak defects at smooth solid areas and of strong defects at the borders of stripes, for $\phi_S\leq0.35$. The later contribution is due to deformations of the contact line at the border of stripes and scales as $\phi_S^2 \ln \phi_S$.
The advancing of the drop represents a rolling motion, and the advancing contact angle was found to be generally anisotropic. The cosine of the longitudinal advancing angle increases linearly on $\phi_S$. The data are well fitted by our simple model, which reflects that the the motion of the contact line at the smooth solid areas is hindered by weak defects. The cosine of the transverse advancing angle is much smaller and has a maximum at  $\phi_S\simeq 0.5$. An explanation of this difference from the longitudinal angle can be obtained if we invoke an additional energy due to strong defects in this direction (governed by the solid area wetted by a drop) which is shown to be proportional to $\phi_S^2$.   Finally, we evaluated the contact angle hysteresis, which was found to be quite large and generally anisotropic. However it becomes practically isotropic in the dilute regime, $\phi_S\leq 0.2$.

Altogether, our study shows that both weak defects at the solid areas and strong defects at the borders of stripes are crucial in determining the contact angle hysteresis. Globally the experimental results are in agrement with the theoretical estimates, confirming that the
our simple approach captures the physical mechanisms at play: although our models may be seen as oversimplified, theoretical results were shown to be predictive over a broad range of solid
surface fraction.

\section*{Acknowledgements}

This work was partly supported by the Russian
Academy of Sciences through its priority program
``Assembly and Investigation of Macromolecular Structures
of New Generations'' and by the
DFG through SFB 985.


\begin{thebibliography}{49}
\bibitem[{App()}]{Apparent}
\bibinfo{note}{We remark that the term apparent contact angle which refers to a
  case, when intrinsic angle is different from what is probed at `macroscale'
  can also be used (see, for example,~\cite{Marmur2009}). However, the notion
  of apparent angle does not necessarily indicate a heterogeneous surface being
  also applied to a homogeneous situation.}
\end{thebibliography}

\end{document}